\documentclass[reprint,amsmath,amssymb]{revtex4-1}

\usepackage{graphicx}
\usepackage{gensymb}
\usepackage{bm}

\begin{document}

\title{Records in Fractal Stochastic Processes}

\author{A. Aliakbari}
\author{P. Manshour}
\email{manshour@pgu.ac.ir}
\author{M. J. Salehi}
\affiliation{Department of Physics, Faculty of Sciences, Persian Gulf University, 75169 Bushehr, Iran}

\begin{abstract}
The records statistics in stationary and non-stationary fractal time series is studied extensively. By calculating various concepts in record dynamics, we find some interesting results. In stationary fractional Gaussian noises, we observe a universal behavior for the whole range of Hurst exponents. However, for non-stationary fractional Brownian motions the record dynamics is crucially dependent on the memory, which plays the role of a non-stationarity index, here. Indeed, the deviation from the results of the stationary case increases by increasing the Hurst exponent in fractional Brownian motions. We demonstrate that the memory governs the dynamics of the records as long as it causes non-stationarity in fractal stochastic processes, otherwise, it has no impact on the records statistics.
\end{abstract}

\maketitle


\section{Introduction}
Emergent extreme events are a key characteristic of complex dynamical systems, and the investigation of such events is a pivotal problem in understanding and predicting various complex systems \cite{Dean:2001aa,Majumdar:2002aa,Burkhardt:2007aa,Sabhapandit:2007aa,Gyorgyi:2008aa}. Recently, the extreme event studies have been developed using various approaches, such as the density of states \cite{Burkhardt:2007aa,Sabhapandit:2007aa}, first-passage and return-time statistics \cite{Majumdar:2001aa,Redner:2001aa,Leadbetter:1982aa}, persistence \cite{Majumdar:1999aa}, interoccurrence time statistics \cite{Manshour:2016aa,Bunde:2003aa,Eichner:2006aa} and record statistics \cite{Chandler:1952aa,Arnold:1998aa,Nevzorov:1988aa}.
Among them, there has been considerable interest in investigating the record statistics, which has found many fruitful applications in diverse complex systems, such as spin glasses \cite{Sibani:1993aa,Sibani:2006aa}, adaptive processes \cite{Orr:2005aa}, domain wall dynamics \cite{Alessandro:1990aa}, avalanche dynamics \cite{Doussal:2009aa}, stock prices \cite{Wergen:2012aa,Wergen:2011aa}, global warming \cite{Wergen:2010aa,Newman:2010aa},  growing network \cite{Godreche:2008aa}, high-temperature superconductors \cite{Oliveira:2005aa}, the ant movements dynamics \cite{Richardson:2010aa}, flood dynamics \cite{WRCR:WRCR9360}, sport statistics \cite{Gembris:2002aa,Gembris:2007aa}, earthquakes \cite{Davidsen:2006aa,Davidsen:2008aa}, and evolutionary biology \cite{Franke:2011aa,Wergen:2013aa}.

A record is an entry in a series that is larger (or smaller) than all previous entries. From another point of view, records are extreme rare events related to an increasing (or decreasing) threshold. The investigation of the record statistics was started by the pioneering work of Chandler \cite{Chandler:1952aa}, which was based on independent and identically distributed (i.i.d.) stochastic time series. One of the most important findings of the record statistics in such random processes is their universal characteristics in the sense that they are completely independent of the underlying distribution \cite{Arnold:1998aa}. To include more complicated systems, many research studies have been performed. For example, the record statistics of independent random variables from non-identical distributions \cite{Krug:2007aa}, broadening distributions \cite{Krug:2007aa,Eliazar:2009aa}, and time-dependent distributions (linear drift) \cite{Wergen:2011ab,Franke:2012aa} are also well understood. However, many time series in real situations are correlated. In this respect, records in symmetric random walks has been well studied, recently \cite{Majumdar:2008aa}, which triggers a series of new research studies, e.g., record statistics of biased random walks and Levy flights \cite{Wergen:2011aa,Wergen:2012aa}, multiple independent random walks \cite{Wergen:2012aa}, and continuous-time random walks \cite{Sabhapandit:2011aa}. The main aim of such studies is to understand and model the statistics of records in observational data by comparing with various kinds of stochastic processes. Despite the striking significance and numerous applications of fractal processes in various areas of science, much less is known about their record statistics. In this regard, it is important to improve our understanding of the record dynamics in such a broader class of stochastic processes. 

In order to model turbulent flows, \textit{fractional Brownian motion} (fBm), a generalization of the more well-known Brownian motion, was introduced many decades ago \cite{Kolmogorov:1940aa,Mandelbrot:1968aa}, and has become one of the most studied stochastic processes widely used in a variety of fields, including physics, probability, statistics, hydrology, economy, biology, and many others \cite{Robinson:2003aa,Embrechts:2002aa,Mandelbrot:2002aa,Alvarez:2006aa,Varotsos:2006aa,Karagiannis:2004aa,Manshour:2015aa}. A fBm is a self-similar Gaussian process with stationary increments (called \textit{fractional Gaussian noise}-fGn) and possesses long-range linear correlation which depends on a parameter, called the Hurst exponent, $H$ \cite{Hurst:1951aa}, where $0<H<1$. The case $H=1/2$ corresponds to the ordinary Brownian motion in which successive increments are statistically independent of one another. For $H>1/2$, the increments of the process are positively correlated (persistent), and for $H<1/2$, consecutive increments are more likely to have opposite signs (anti-persistent). 

The rest of this paper is organized as follows. In section~\ref{GCRD}, we review and discuss some basic definitions and general findings in the record theory. In particular, we present important findings from two most studied time series: an i.i.d. process and a symmetric random walk. Next, by using extensive numerical analysis, we study the record statistics of non-stationary fractional Brownian motions as well as stationary fractional Gaussian noises in section~\ref{RF}. We show that for stationary fractal series a robust universal behavior in the record dynamics is observed. We also demonstrate that the non-stationarity in fractal processes destroys such a universality observed in the stationary case. Finally, we conclude in section~\ref{Conc}.

\section{General concepts in records statistics}
\label{GCRD}

\begin{figure}[t]
\begin{center}
\includegraphics[scale=0.5]{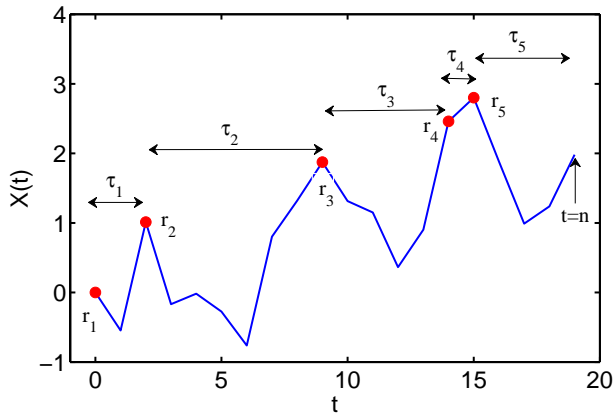}
\caption{A realization of a symmetric random walk $X(t)$ with $n$ steps. Here the number of records is $R_n=5$. The $\tau_i$'s denote the ages of the records and the $r_i$'s are the records values.}
\label{records}
\end{center}
\end{figure}

In a series of random variables $X_1, X_2, ..., X_n$, the data point $X_n$ is a record if $X_n>\text{max}\{X_0,X_1,...,X_{n-1}\}$. Fig.~\ref{records} represents the records evolution in a symmetric random walk, in which filled red dots show the records. One of the fundamental quantities in the records theory is the record rate, $P_n$, defined as the probability that $X_n$ is a record. To simplify the analysis, it is helpful to introduce a quantity called the record indicator, as follows:
\begin{equation}
  I_n=\left\{
  \begin{array}{@{}ll@{}}
    1, & \text{if}\ X_n \text{ is a record,} \\
    0, & \text{otherwise}
  \end{array}\right.
\end{equation}
Then, in terms of the record indicator, we can write
\begin{equation}
P_n = \text{Prob} [I_n=1]
\end{equation}
Hence, the ensemble averaging of the record indicator $\bar{I}_n$ at time step $n$, reads, 
\begin{equation}
\label{I_P}
\bar{I}_n = P_n \times 1 + (1-P_n) \times 0=P_n
\end{equation}
In practice, $\bar{I}_n$ represents the value that one would obtain after collecting and averaging many samples of the time series. The number of records $R_n$ that occur up to time $n$ is another quantity of particular interest (see Fig.~\ref{records}), and defined as the sum over the record indicator
\begin{equation}
R_n=\sum_{i=1}^n{I_i}
\end{equation}
This means that the average record number $\overline{R_n}$ is just the sum over the record rate $P_i$,
\begin{equation}
\label{R_avg}
\overline{R_n}=\sum_{i=1}^n{P_i}
\end{equation}

For i.i.d. random time series, due to the equal probability of the record occurrence at each time step, one finds that $P_n=1/n$, which is completely independent of the underlying distribution of the series \cite{Nevzorov:1988aa,Arnold:1998aa}. By substituting this into Eq.~\ref{R_avg}, one obtains for the average record number $\overline{R_n} \simeq \ln{n}+\gamma$, where $\gamma \simeq 0.577215...$ is the Euler-Mascheroni constant \cite{Abramowitz:1970aa}. The mean number of records is an increasing (but extremely slow) function of time, due to the fact that the probability of the record occurrence decays with time. Also, the distribution of the record number, $P(R_n)$, over a large number of realizations of i.i.d series approaches a Gaussian function, with mean and variance of $\ln{n}$ \cite{Krug:2007aa}, as follows:
\begin{equation}
\label{P_R_iid}
P(R_n)=\frac{1}{\sqrt{2\pi \ln{n}}}e^{-\frac{(R_n-\ln{n})^2}{2\ln{n}}}
\end{equation}

Similar universal properties have been found for the record statistics of symmetric random walks, by  Majumdar and Ziff \cite{Majumdar:2008aa}. They showed that for a symmetric random walk $\overline{R_n} \simeq \sqrt{4n/\pi}$ and $P_n\simeq 1/\sqrt{\pi n}$ as $n \rightarrow \infty$. As it can be seen, the record rate $P_n$ of symmetric random walks decays much slower than that of i.i.d. random series. They also computed the record number distribution, and showed that it approaches a half-Gaussian form of
\begin{equation}
\label{P_R_rw}
P(R_n)=\frac{1}{\sqrt{\pi n}}e^{-\frac{R_n^2}{4n}}
\end{equation}
where the most probable value of $R_n$ is zero. We note that the standard deviation of $R_n$ is large ($\sigma_{R_n}\sim \sqrt{n}$) for large $n$, in contrast to the i.i.d. case which is small ($\sigma_{R_n}\sim \sqrt{\ln{n}}$) compared to the mean. This indicates that the record number distribution of a (correlated) symmetric random walk is significantly broader than that of an (uncorrelated) i.i.d. time series.

In order to search for the time characteristics of the record statistics, one may take into account the record ages. Let $\tau=[\tau_1, \tau_2, ..., \tau_R]$ denote the time intervals between successive records. On the other word, $\tau_i$ is the age of the $i$th record; i.e., it denotes the time up to which the $i$th record survives (see Fig.~\ref{records}). Note that the last record, i.e., the $R_n$th record, still stays a record at the $n$th step since there are no more record breaking events after. The typical age of a record can be estimated as $\tau_{typ} = n/\overline{R_n}$, which behaves as $n/\ln{n}$ and $\sqrt{\pi n/4}$ for i.i.d. and random walk time series, respectively \cite{Majumdar:2008aa}. However, there are rare records with completely different age behavior. For example, one may consider the longest age $\tau_{max}=\text{max}[\tau_1,\tau_2,...,\tau_{R_n}]$ of the records. In particular, the asymptotic large $n$ behavior of the average of $\tau_{max}$ can be extracted explicitly as $\bar{\tau}_{max} \simeq cn$, with $c\simeq0.624330$ and $0.626508$ for i.i.d. and symmetric random walk series, respectively \cite{Majumdar:2008aa}. This shows that the longest age of the records is much larger than the typical one. 

Further, one may ask wether the record occurrences are correlated. To quantify such a correlation, we can take into account the joint probability $P_{n,m}=\text{Prob} [I_n=1 \text{ and } I_m=1]$ of the occurrence of two records at time steps $n$ and $m$. It was shown \cite{Arnold:1998aa} that for i.i.d. time series, the individual record events are uncorrelated, and our knowledge about the present record tells us nothing about the future one. This means that the record events of $n$ and $m$ are completely independent if $m\neq n$, and the joint probability becomes $P_{n,m}=P_nP_m$. In the language of probability theory, for the conditional probability we have $P_{n|m}=P_{n,m}/P_m$, which defines the occurrence probability of a record at time step $n$, given that a record has occurred in time step $m$. Thus, in the case of i.i.d. processes, $P_{n|m}$ is simply equal to $P_n$. For a symmetric random walk, records are not independent, and the conditional probability of two successive records is given by $P_{n+1|n}=1/2$ for large $n$ \cite{Wergen:2011ab}. This shows that the record events in symmetric random walks effectively attract each other, i.e., there is an increased probability for the occurrence of a record if another record has occurred in the previous time step \cite{Wergen:2011ab}.

\section{Records in fractals}
\label{RF}

In this section, we present an extensive numerical analysis of the record statistics for fractal time series with various Hurst exponents. We discuss the effect of correlation, as well as stationarity/non-stationarity on various aspects of the record dynamics. In order to investigate the record statistics of fractional processes, we construct such time series through a generic $1/f$ noise with Fourier filtering method \cite{Makse:1996aa}. Then, we search for various record features of fractional time series of size up to $N=10^5$. Also, all calculations have been performed by averaging over $10^6$ different realizations.

\subsection{Record rate}
\label{Rate}

\begin{figure}[t]
\begin{center}
\includegraphics[scale=0.4]{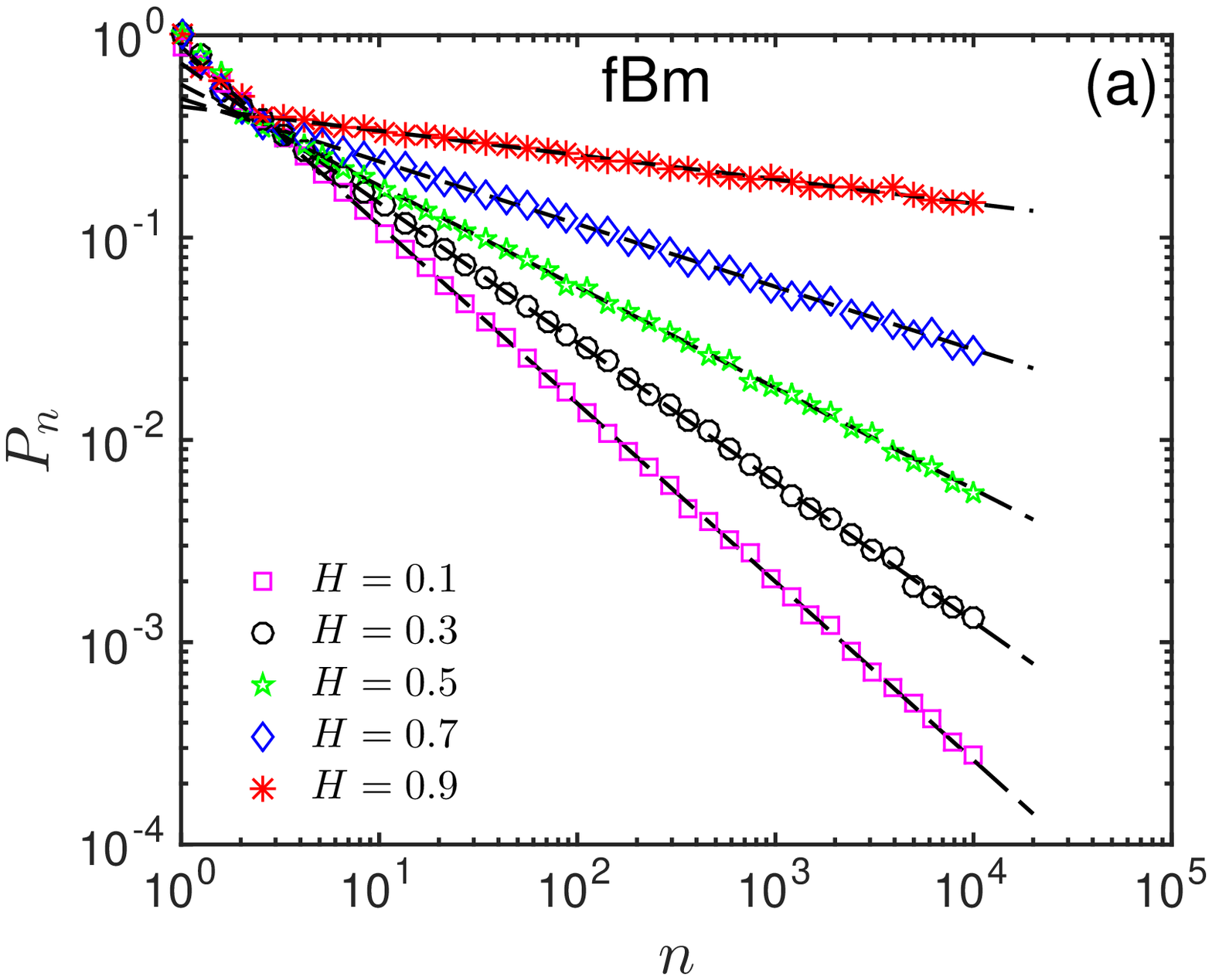}
\includegraphics[scale=0.4]{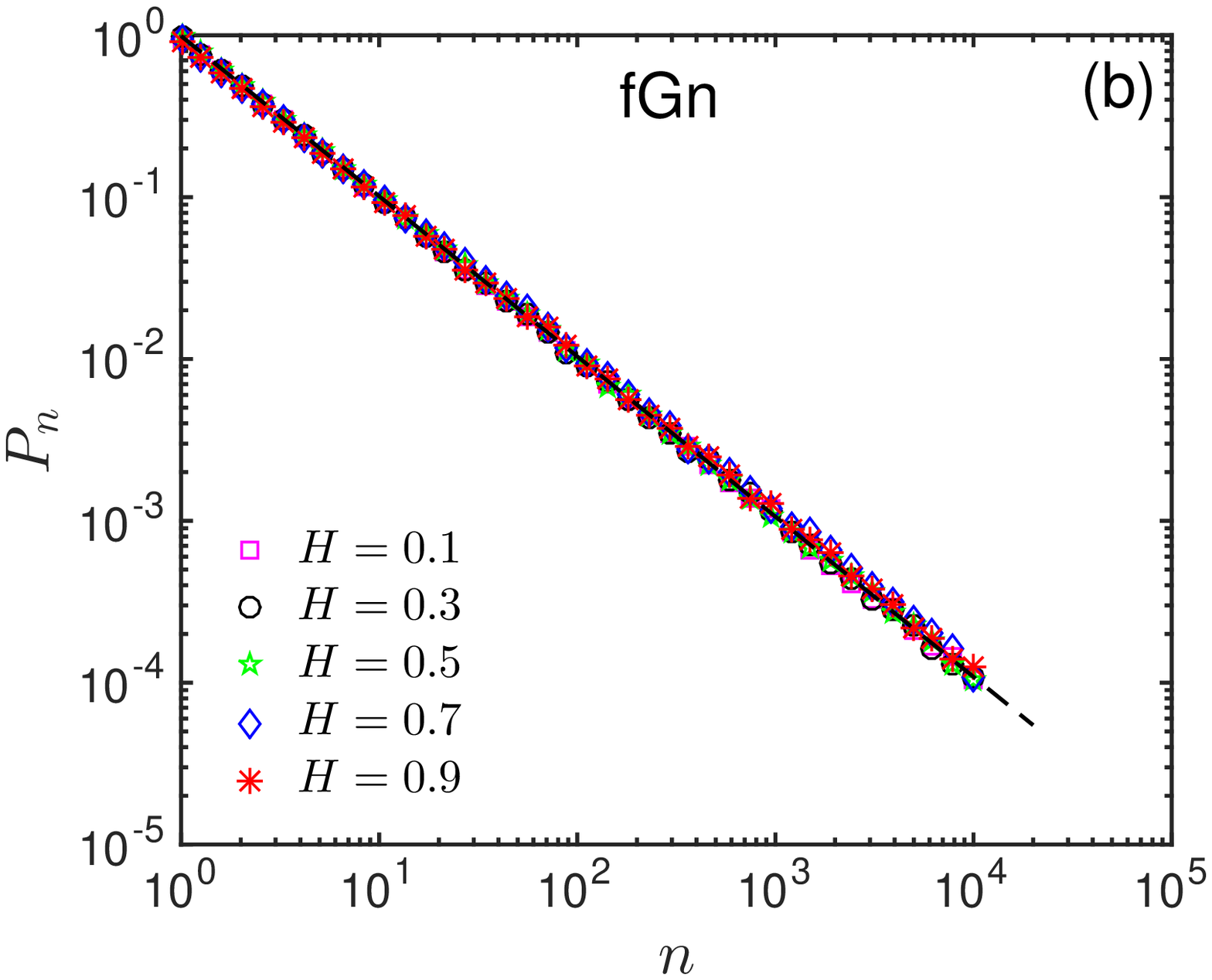}
\caption{The record rates $P_n$ for (a) fBm, and (b) fGn series with five different Hurst exponents. The dashed lines represent the fitted power-law functions, discussed in section~\ref{Rate}.}
\label{P_n}
\end{center}
\end{figure}

At first, we construct series of record events, corresponding to fractional Brownian motions and fractional Gaussian noises with different Hurst exponents, and search for various statistics. In Fig.~\ref{P_n}(a), we plotted in logarithmic scale, the record rates $P_n$ for fBm series with different Hurst exponents of $0.1$, $0.3$, $0.5$, $0.7$, and $0.9$. We fit power-law functions (dashed lines) of the form of $P_n \sim n^{-\delta}$ to the record rates, and find that $\delta=1-H$. Majumdar and Ziff showed \cite{Majumdar:2008aa} that the probability $P_n$ for a record in the $n$th event of a symmetric random walk is the same as the corresponding persistence probability $Q_n=\text{Prob}[X_1, X_2, ..., X_n <0]$, which is the probability that a random walker starting from the origin stays bellow the origin without crossing it for the next $n$ steps, and given by $Q_n=P_n=1/\sqrt{\pi n}$ as $n \rightarrow \infty$. On the other hand, it was shown \cite{Krug:1997aa} that the persistence probability for a fBm series is given by $Q_n\simeq n^{H-1}$. Here, we find numerically such a power-law behavior for the record rate of fractional Brownian motions.

We also search for the record rates corresponding to fractional Gaussian noises. Fig.~\ref{P_n}(b), shows log-log plots of the record rates, $P_n$, for different Hurst exponents. Interestingly, all the curves collapse into a universal power-law function of the form of $P_n\sim 1/n$ (dashed line) for the whole range of Hurst exponents. This result is the same as the one obtained for an i.i.d. random series. Indeed, we find that the record rate in stationary fractional Gaussian noises is independent of the underlying memory in the time series. We will discuss this finding in more details in section~\ref{cor_rec}, by searching for correlations between the records.

\subsection{Record number}
\label{rec_num}

\begin{figure}[t]
\begin{center}
\includegraphics[scale=0.4]{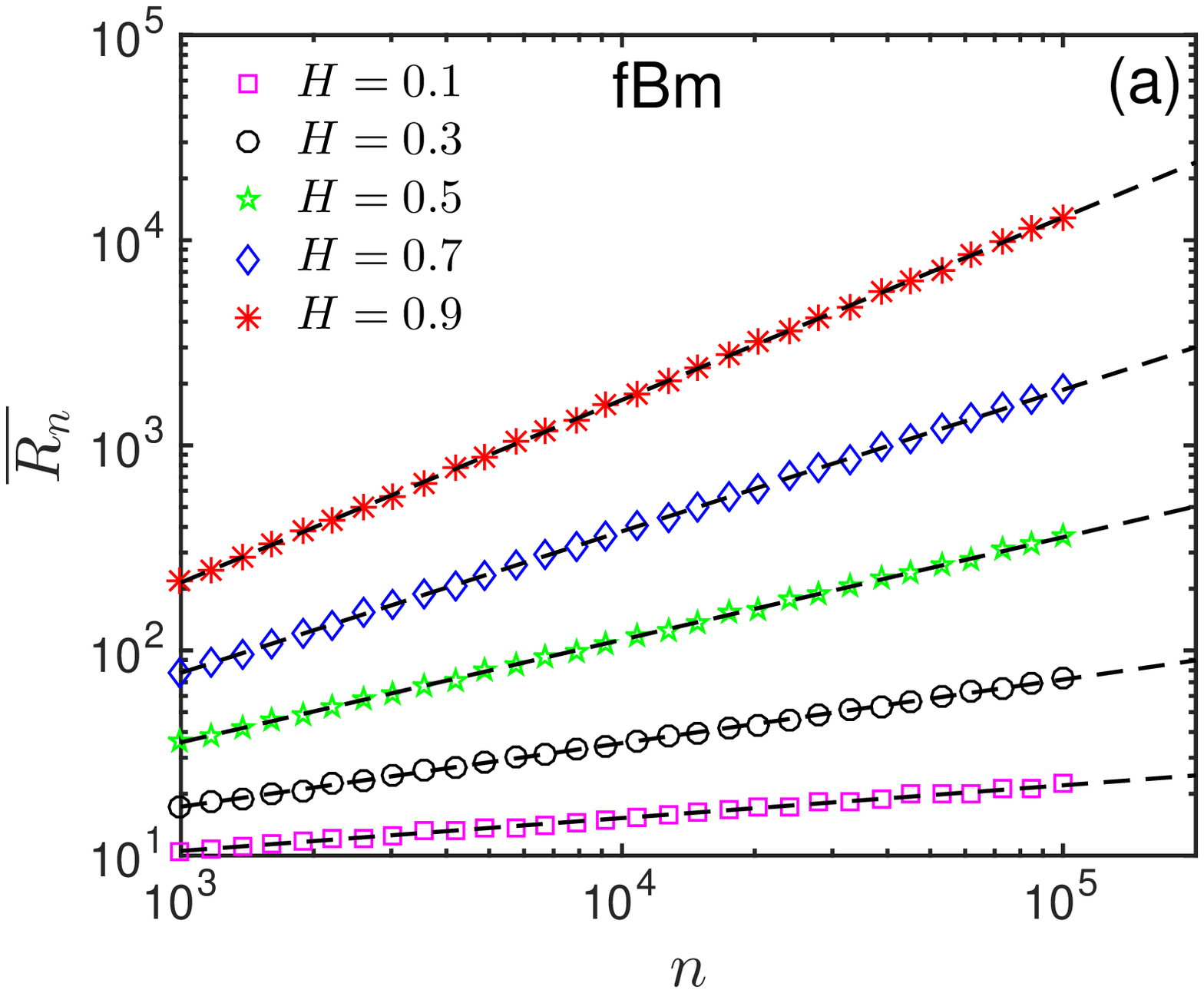}
\includegraphics[scale=0.4]{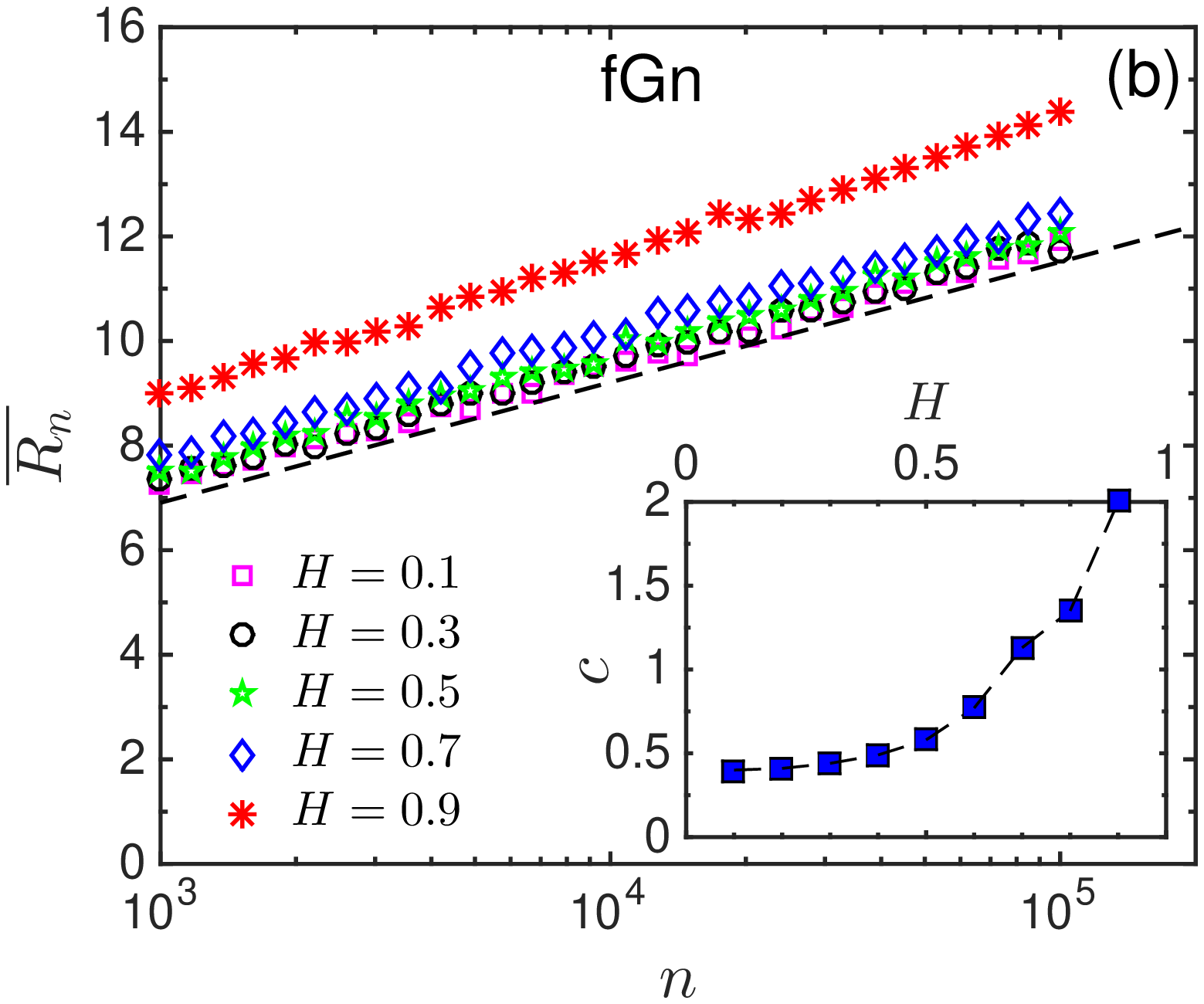}
\caption{The time behavior of the average record number $\overline{R_n}$ for (a) fBm, and (b) fGn series with five different Hurst exponents. The dashed lines in (a) represent the fitted power-law functions of $n^H$, discussed in section~\ref{rec_num}. The inset in (b) shows the fitting parameter $c$ in $\overline{R_n}\simeq \ln{n}+c$ for different Hurst exponents. Also, the dashed line in (b) shows a $\ln{n}$ function, for comparison.}
\label{Rn_avg}
\end{center}
\end{figure}

\begin{figure}[t]
\begin{center}
\includegraphics[scale=0.4]{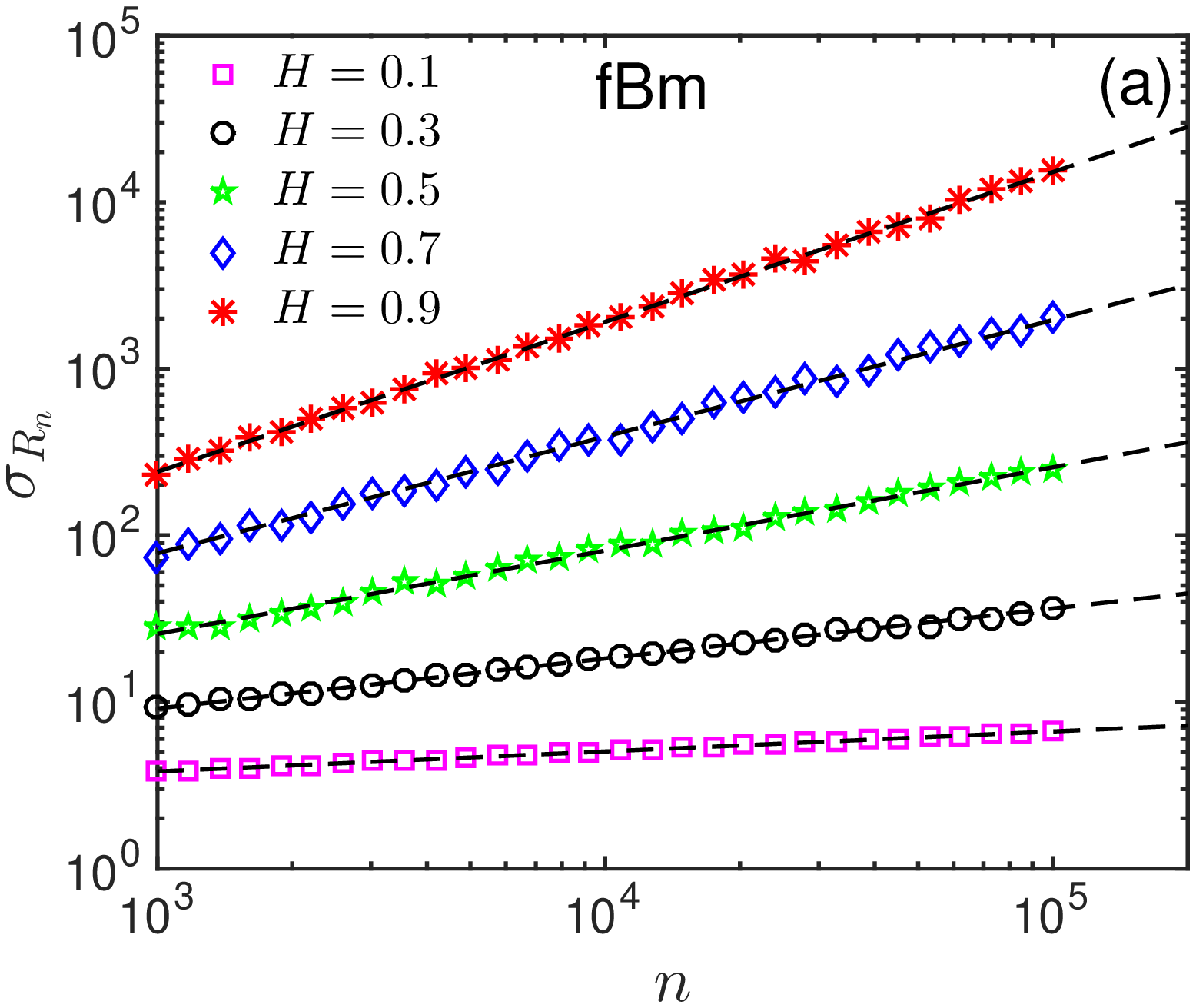}
\includegraphics[scale=0.4]{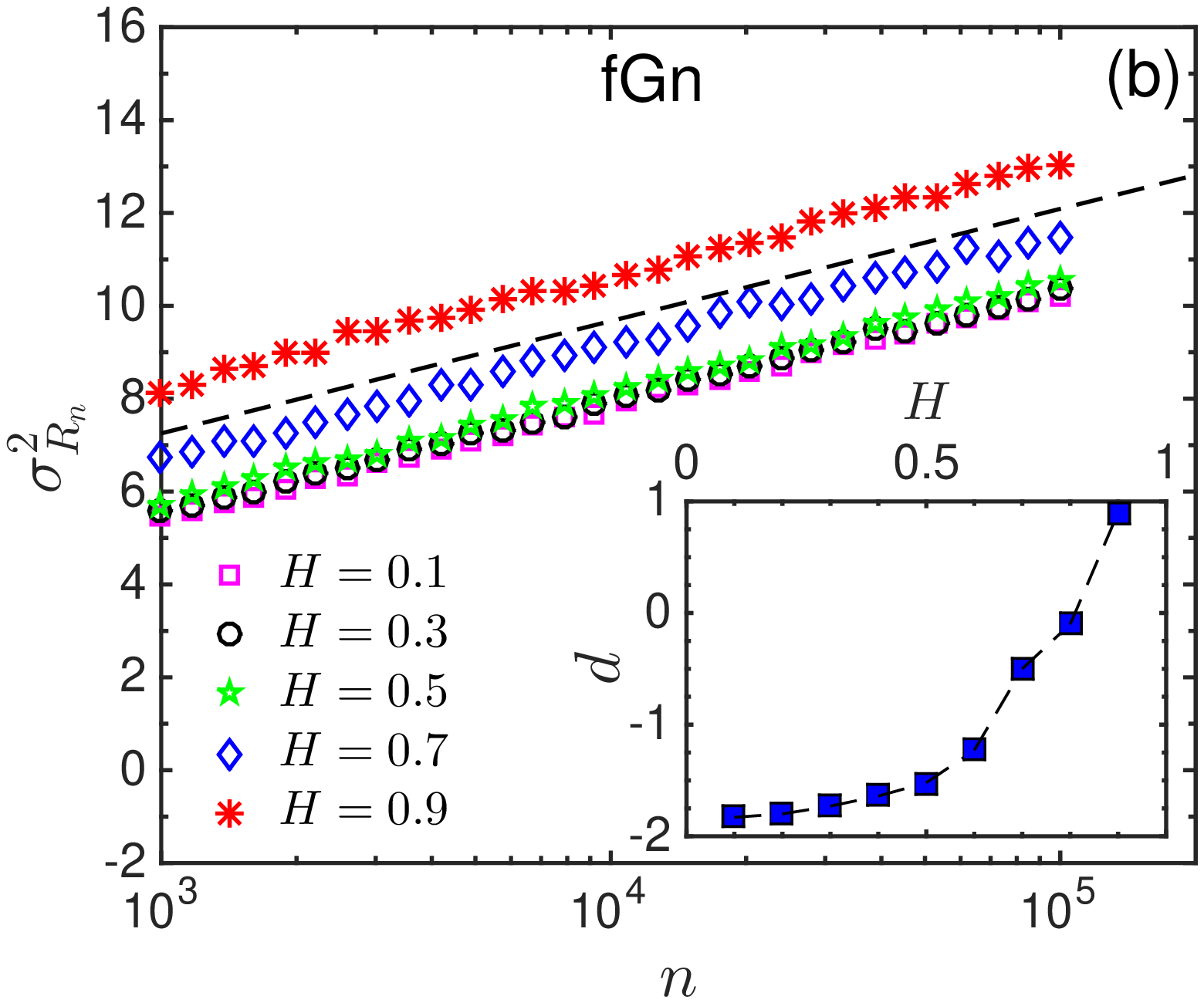}
\caption{The time behavior of the standard deviation $\sigma_{R_n}$ for (a) fBm, and (b) fGn series with five different Hurst exponents. The dashed lines in (a) represent the fitted power-law functions of $n^H$, discussed in section~\ref{rec_num}. The inset in (b) exhibit the fitting parameter $d$ in $\sigma^2_{R_n}\simeq \ln{n}+d$ for different Hurst exponents. Also, the dashed line in (b) shows a $\ln{n}$ function, for comparison.}
\label{Rn_std}
\end{center}
\end{figure}

As we mentioned above, the record number is an important quantity in the record theory. Fig.~\ref{Rn_avg}(a) demonstrates the average record number $\overline{R_n}$ versus time for fBm series with five different Hurst exponents. We find a power-law behavior (dashed lines) of the form of $\overline{R_n}\sim n^{H}$. However, a completely different behavior is observed for fGn series. Fig.~\ref{Rn_avg}(b) shows semi-logarithmic plots of $\overline{R_n}$ for fGn series with various Hurst exponents. We find a logarithmic increase as $\overline{R_n}\simeq \ln{n}+c$, where $c$ depends on the Hurst exponent, as indicated in the inset of Fig.~\ref{Rn_avg}(b). Clearly, the mean record number for a non-stationary fBm series grows much faster than that for a stationary fGn series. Further, we also search for the fluctuations about these average values. In Fig.~\ref{Rn_std}, we plotted the standard deviation of the record number against time for fBm and fGn series with different Hurst exponents. We find again a power-law form of $\sigma_{R_n}\sim n^{H}$ for fBm, and a logarithmic form of $\sigma^2_{R_n}\simeq \ln{n}+d$ for fGn series. Here, $d$ is also dependent on the Hurst exponent, as shown in the inset of Fig.~\ref{Rn_std}(b).

\begin{figure}[t]
\begin{center}
\includegraphics[scale=0.4]{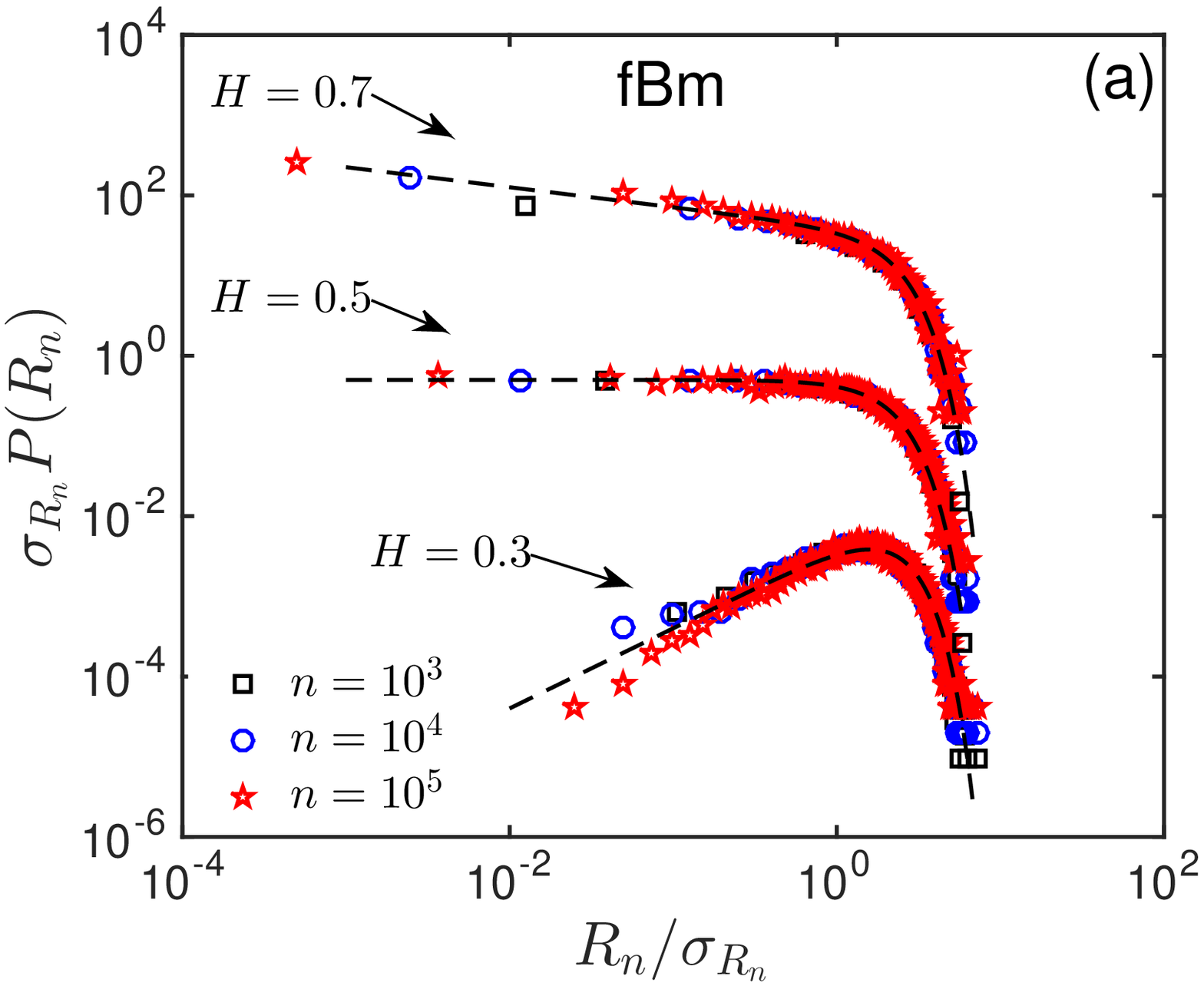}
\includegraphics[scale=0.4]{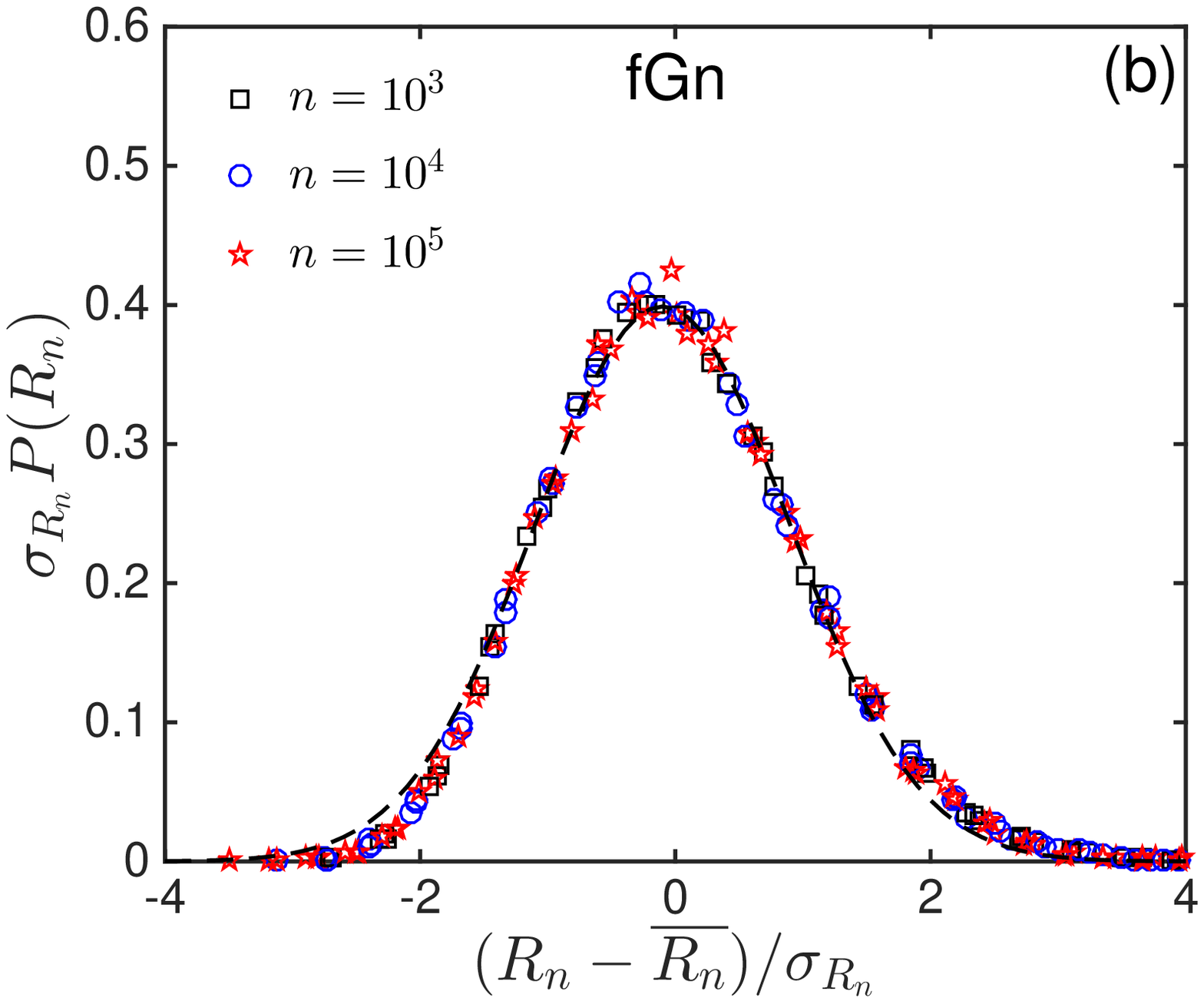}
\caption{The normalized record number distributions, $P(R_n)$, for (a) fBm, and (b) fGn series with three different Hurst exponents of $0.3$, $0.5$, and $0.7$ and three different sizes. In (a), the plots for $H=0.3$ and $H=0.7$ are shifted vertically by a factor of $0.01$ and $100$ for better visibility. In (b) all curves with different Hurst exponents collapse into a universal Gaussian function. The dashed lines in both figures represent the corresponding fitted functions, discussed in section~\ref{rec_num}.}
\label{P(Rn)}
\end{center}
\end{figure}

To find out more about the nature of the record number fluctuations, it is helpful to take into account the record number distributions $P(R_n)$. In Fig.~\ref{P(Rn)}(a), we plotted the normalized probability distributions of the record number for fBm series of sizes $n=10^3$, $10^4$, and $10^5$, and with three different Hurst exponents of $0.3$, $0.5$, and $0.7$. Due to better visibility, the curves in Fig.~\ref{P(Rn)}(a) are shifted vertically by a factor of $0.01$ and $100$ for $H = 0.3$ and $H = 0.7$, respectively. The dashed lines show fitted functions of the form of 
\begin{equation}
\label{P_R_fbm}
y\sim x^{\alpha} e^{-\frac{x^2}{5}}
\end{equation}
where $x=R_n/\sigma_{R_n}$, $y=\sigma_{R_n}P(R_n)$, and $\alpha$ is positive/zero/negative for Hurst exponents smaller than/equal to/larger than $0.5$. We observe that $P(R_n)$ approaches an asymmetric Gaussian form, as $H\rightarrow 0$, and tends to a power-law functional form, as $H\rightarrow 1$. Therefore, in non-stationary fractal processes, increasing $H$ broadens the record number distribution. Note that for $H=0.5$, we find the half-Gaussian distribution of Eq.~\ref{P_R_rw}, as expected.

Fig.~\ref{P(Rn)}(b) shows the record number distributions for stationary fGn series, with different sizes and Hurst exponents. The curves are normalized to zero mean and unit standard deviation. As it can be seen, all curves collapse into a universal Gaussian function of the form of
\begin{equation}
\label{P_R_fgn}
y=\frac{1}{\sqrt{2\pi}} e^{-\frac{x^2}{2}}
\end{equation}
where $x=(R_n-\overline{R_n})/\sigma_{R_n}$ and $y=\sigma_{R_n}P(R_n)$. This indicates that the record number distribution of stationary fractal time series is the same as that of an i.i.d random series of Eq.~\ref{P_R_iid}, and is completely independent of the memory aspects of the original process. 
We note here that for non-stationary fractal series, the width of the record number distribution increases as the Hurst exponent increases, in contrast with the case of stationary fGn series. This indicates that for high values of the Hurst exponent in an fBm series, the record parameters can largely fluctuate about their mean values, and thus we should be careful when we talk about such average values. 

\subsection{Record ages}
\label{ages}
\begin{figure}[t]
\begin{center}
\includegraphics[scale=0.4]{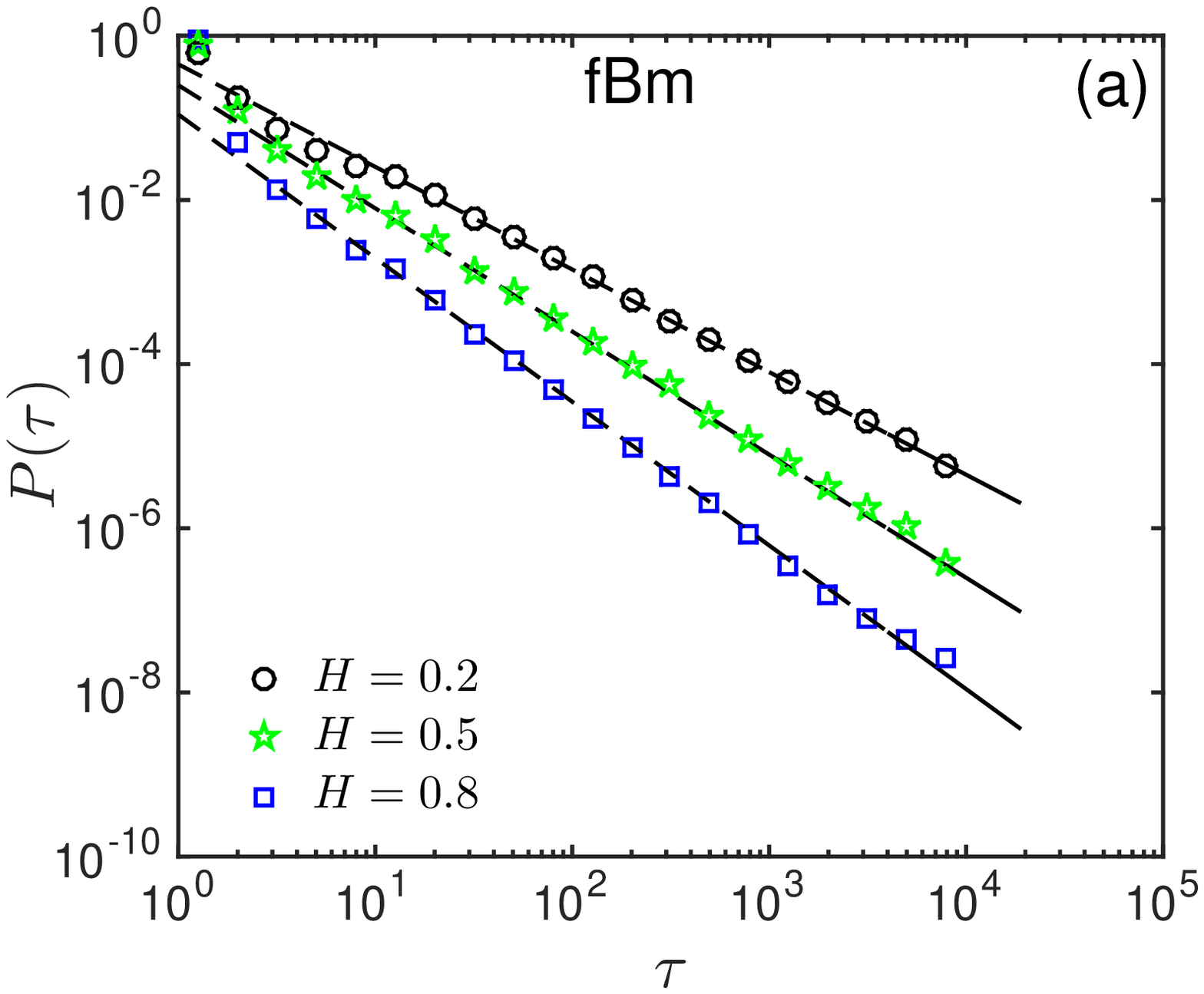}
\includegraphics[scale=0.4]{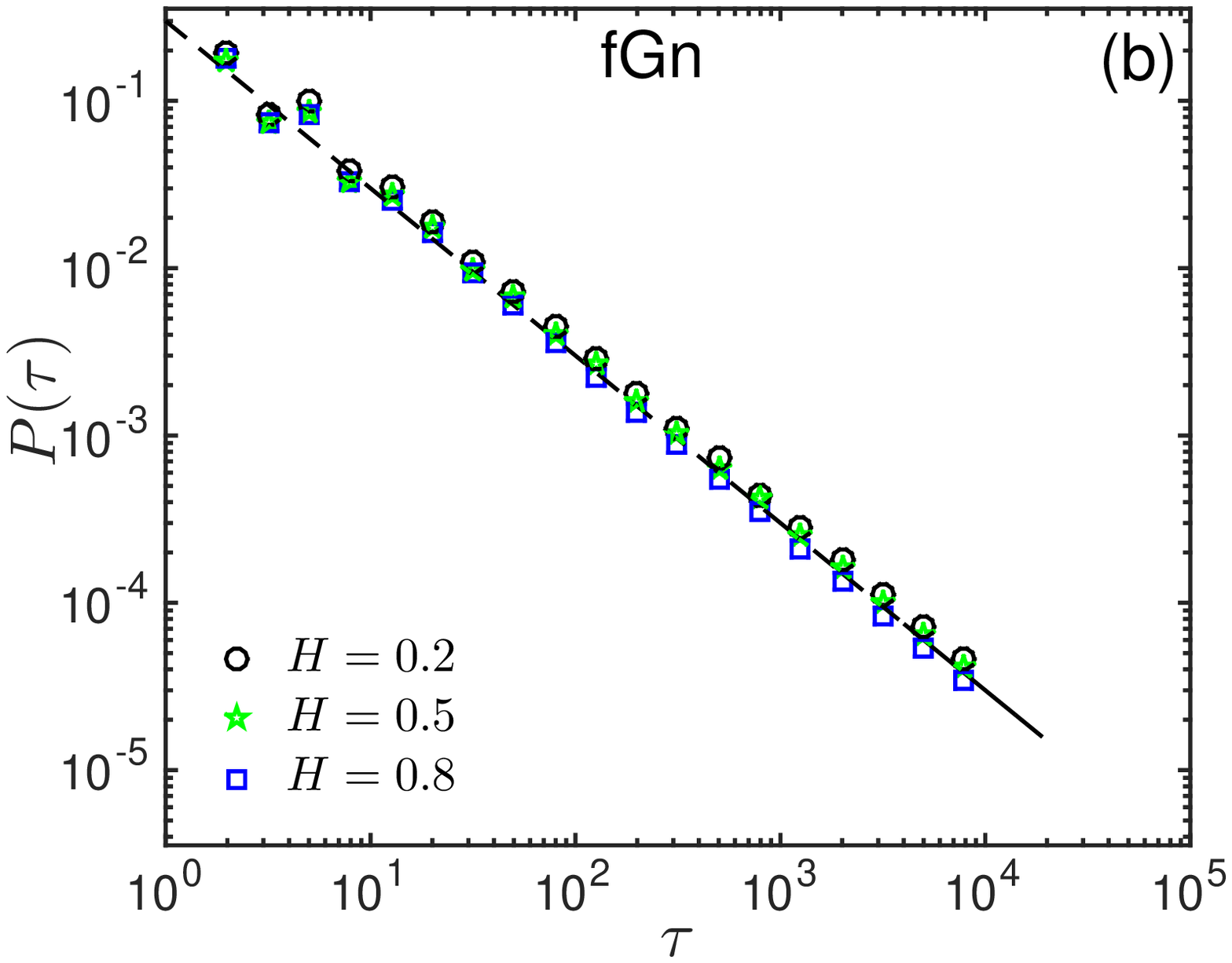}
\caption{The age distributions $P(\tau)$ of the records for (a) fBm, and (b) fGn series of size $n=10^5$ with three different Hurst exponents of $0.2$, $0.5$, and $0.8$. The dashed lines in both figures represent the corresponding fitted power-law functions, discussed in section~\ref{ages}.}
\label{P_T}
\end{center}
\end{figure}

We are also interested in questions like ``how long should one wait for the occurrence of the next record?''. In this respect, we calculate the age distribution $P(\tau)$ of the records for both fGn and fBm series with various Hurst exponents. Fig.~\ref{P_T}(a) represents such distributions for fBm series with three Hurst exponents of $0.2$, $0.5$, and $0.8$. The dashed lines exhibit power-laws of the form of
\begin{equation}
P(\tau)\sim \tau^{-\beta}
\end{equation}
with $\beta=1+H$. In the case of fGn series, we find again a universal power-law behavior of the form of $P(\tau)\sim \tau^{-1}$ (dashed line in Fig.~\ref{P_T}(b)) for different Hurst exponents. It is interesting to note here that any questions about the average of the record age does not make sense, since due to the power-law behavior of the age distribution with exponents between $-2$ and $-1$, $\overline{\tau}=\int{\tau P(\tau)d\tau}$ diverges as $\tau\rightarrow \infty$.

\subsection{Correlation in records}

\begin{figure}[t]
\begin{center}
\includegraphics[scale=0.4]{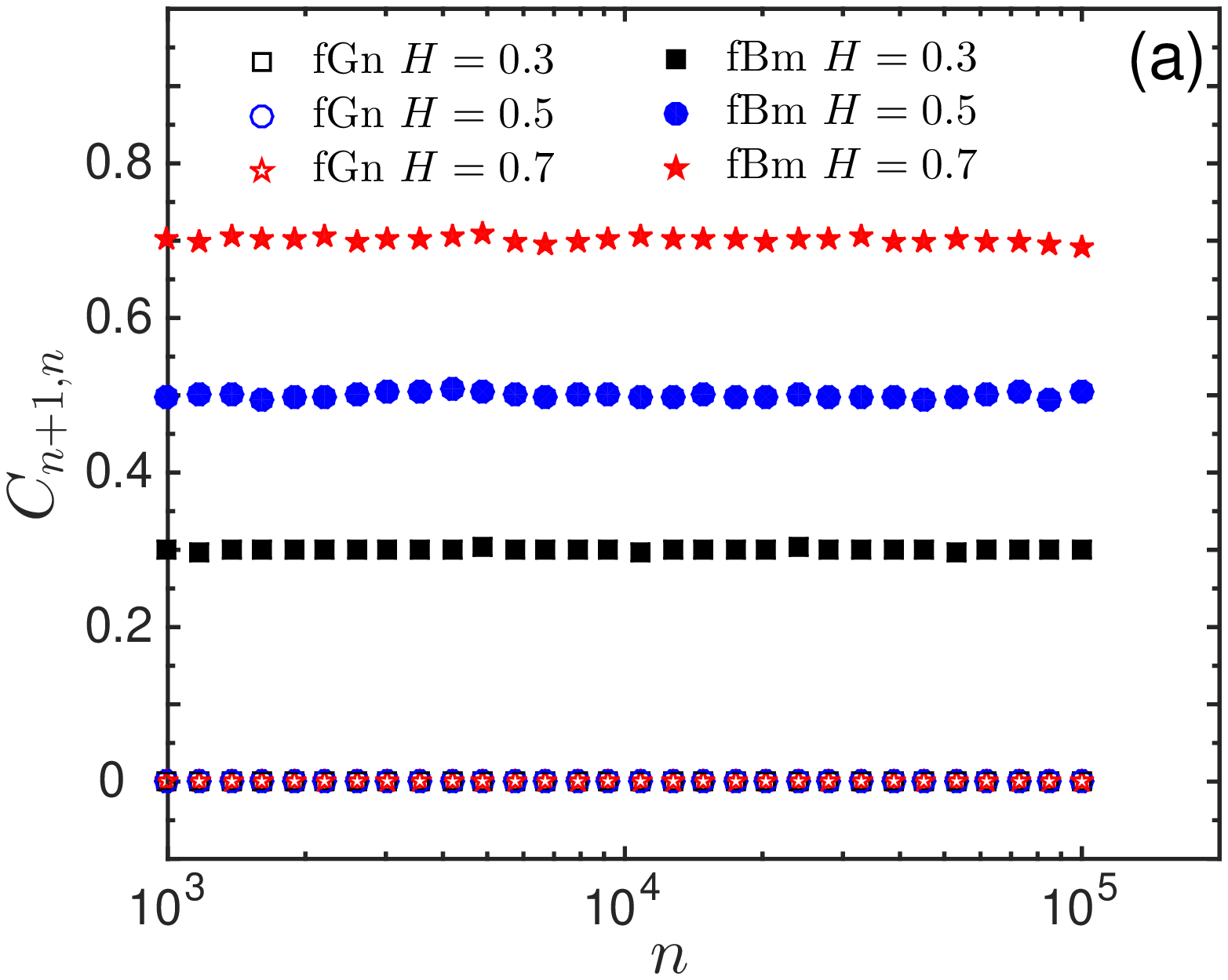}
\includegraphics[scale=0.4]{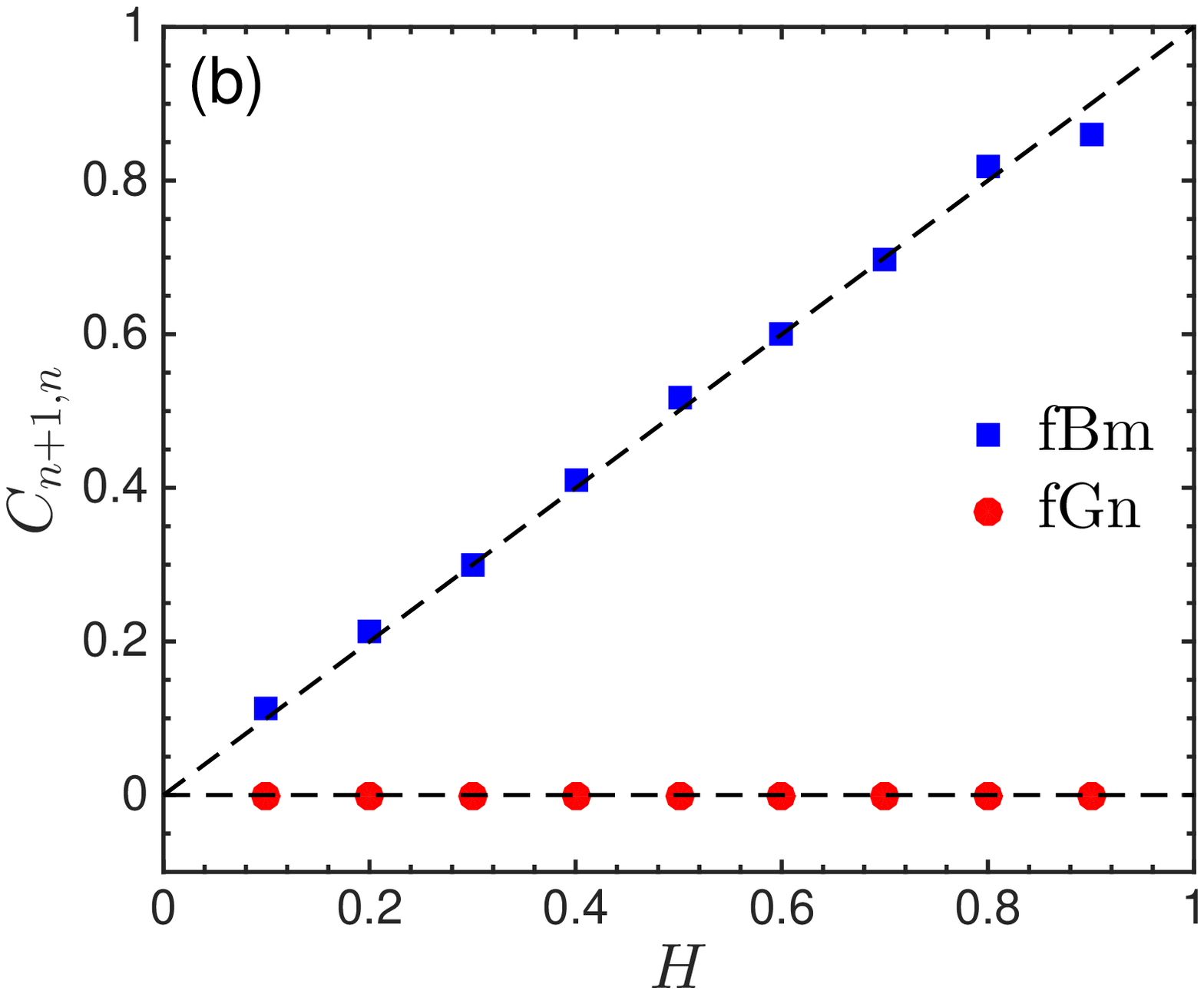}
\caption{(a) the time behavior and (b) the Hurst dependency of the correlation index $C_{n+1,n}$, as defined in Eq.~\ref{C_n}, for fBm and fGn time series. The dashed lines in (b) show the fitted functions, as discussed in section~\ref{cor_rec}. As it can be seen, the records occurrence in stationary fGn series are uncorrelated.}
\label{Corr}
\end{center}
\end{figure}

\label{cor_rec}
At the end, we are interested in correlations between the individual record events. To quantify the correlation properties between two records at time steps $n$ and $m$, we introduce a correlation index, $C_{n,m}$, as follows:
\begin{equation}
C_{n,m}=\frac{\overline{I_nI_m}-\bar{I}_n\bar{I}_m}{\sigma_{I_n}\sigma_{I_m}}
\end{equation}
where $\sigma_{I}^2=\overline{I^2}-\overline{I}^2$. Due to the binary property of the record indicator, $I^2=I$, and one simply gets $\sigma_{I}^2=\overline{I}-\overline{I}^2$. On the other hand, the joint probability $P_{n,m}$ of the occurrence of two record events can be written, in terms of the record indicator $I_n$, as
\begin{equation}
P_{n,m}=\text{Prob} [I_n=1 \text{ and } I_m=1]=\overline{I_nI_m}
\end{equation}
Then, by using Eq.~\ref{I_P}, we have
\begin{equation}
C_{n,m}=\frac{P_{n,m}-P_nP_m}{\sqrt{P_n-P^2_n}\sqrt{P_m-P^2_m}}
\end{equation}
Hence, for an independent sequence with $P_{n,m}=P_nP_m$, we obtain $C_{n,m}=0$. From now on, we concentrate our attention only on the correlation between two successive points, i.e., $C_{n+1,n}$. According to the power-law form of the record rate, we get $P_{n+1}/P_n=(1+1/n)^{-\delta}$. Thus, one can assume that $P_{n+1} \simeq P_{n}$ for $n\gg 1$. By using the relationship $P_{n+1,n}=P_{n+1|n}P_n$, we have
\begin{equation}
\label{C_n}
C_{n+1,n}\simeq \frac{P_{n+1|n}P_n-P_n^2}{P_n-P_n^2}
\end{equation}
For $n\gg 1$, we can also neglect $P_n^2$ in comparison with $P_n$, and then finally we get $C_{n+1,n}\simeq P_{n+1|n}$. We know that for a fBm process, the conditional probability $P_{n+1|n}$ equals to the probability of the occurrence of a positive increment, and equals to $H$. Thus, we find $C_{n+1,n}\simeq H$ which is independent of $n$. To check this, we plotted in Fig.~\ref{Corr}(a) the time behavior of the correlation index, $C_{n+1,n}$, for fractal time series with different Hurst exponents. We observe that for fBm series, $C_{n+1,n}\simeq H$, and is approximately independent of time for $n\gg 1$, as expected. Also, for stationary fGn series, we find $C_{n+1,n}\simeq 0$, independent of the Hurst exponent. This indicates that the record events in stationary fractal time series are uncorrelated for large $n$. For better understanding, we also plotted in Fig.~\ref{Corr}(b), the correlation index $C_{n+1,n}$ at a fixed time step $n=5000$, for various Hurst exponents. As it can be seen, we find a linear behavior of $C_{n+1,n}$ versus $H$ for fBm series, in perfect agreement with our analytical result. Interestingly, for fGn series with various Hurst exponents, the record dynamics is uncorrelated and independent of the Hurst exponent. We find an important result: the record statistics of stationary fractal time series are uncorrelated, even with the presence of long-range linear correlations in the original process. In fact, this feature influences all aspects of the record dynamics, causing the observed universal behavior in the records statistics, mentioned above.

\section{conclusion}
\label{Conc}

In conclusion, we investigated the record statistics of stationary and non-stationary fractal time series, i.e., fractional Gaussian noises and fractional Brownian motions, respectively. By doing extensive numerical calculations, we found a power-law functional form for many record features such as the record rates and the age distributions. We have also investigated other dynamical characteristics of the records, such as correlation between the record events, and found that the record dynamics are uncorrelated in stationary fractal time series, in spite of the existence of long-range linear correlation in the original series. Further, we have shown that for stationary fractal time series, a universal behavior is observed in all aspects of the record dynamics, independent of the memory inherited in the process. On the other hand, for non-stationary fractal processes, the record dynamics strongly depends on the Hurst exponent. This deviation from the results of the stationary case is increased by increasing the Hurst exponent, which can be considered as the degree of non-stationarity in such fractal series. Finally, we have demonstrated that the memory in fractal time series can affect the record dynamics, only if the existing correlation can produce non-stationarity in such fractal processes.
\begin{acknowledgments}
The support from Persian Gulf University Research Council is kindly acknowledged.
\end{acknowledgments}

\bibliography{records}

\end{document}